\documentclass[twocolumn,aps,prl,english,twocolumn,epsfig,groupedaddress]{revtex4}
\usepackage{graphicx}
\usepackage[latin1]{inputenc}
\usepackage{latexsym}

\begin{document}
\title{\bf Magnetic spin excitations in diluted ferromagnetic systems: the case of Ga$_{1-x}$Mn$_{x}$As.}

\author{Georges ~Bouzerar$^{1,2}$\footnote[5]{email:bouzerar@ill.fr}}

\affiliation{ 
$^{1}$ Institut N\'eel, MCBT, CNRS, 25 avenue des Martyrs, B.P. 166 38042 Grenoble Cedex 09
France.\\
$^{2}$ Institut Laue Langevin BP 156 38042 Grenoble Cedex
France.\\
}            
\date{\today}

\begin{abstract}
\parbox{14cm}{\rm}
\medskip
By using an approach based on the Self-Consistent Local Random Phase Approximation (SC-LRPA) we calculate the magnetic excitation spectrum in diluted/disordered ferromagnetic systems. In previous studies, the SC-LRPA was shown to be reliable for the determination of the Curie temperature. In this letter, we now demonstrate its accuracy and efficiency for the determination of the magnetic excitation spectrum. For that purpose, we calculate the spin excitations of the widely studied diluted III-V magnetic semiconductor Ga$_{1-x}$Mn$_{x}$As. Instead of the expected broadening of the excitations due to disorder, it is shown that magnons exist only in a very restricted region of the Brillouin zone. We calculate without adjusting parameters, the spin stiffness in {\it optimally annealed} systems as a function of $x$ and compare it to recent measurements. We find a very good agreement for well annealed samples and provide a plausible explanation for the very small values measured in as grown ones. We hope that this study will stimulate new studies as Ineslatic Neutron Scattering measurements for example.
\end{abstract}
\pacs{PACS numbers:  75.10.-b, 75.50.Pp, 75.40.Gb}
\maketitle

The understanding of the influence and effects of disorder on transport and magnetic properties in magnetic materials is a key issue from both a fundamental point of view or in the prospect of possible technological applications. This requires the necessity to develop new methods and tools which are able to treat the disorder in a reliable manner. The word disorder is very general and includes for example,(i) the dilution of magnetic atoms in a non magnetic matrix, (ii) substitution of a non magnetic atom by another one which has a different radius or/and valency as in manganites A$_{1-x}$B$_{x}$MnO$_3$ where A$=$La,Pr and B$=$Sr,Ca,Ba for example, (iii) intrinsic defects which may appear during the growth of the material, for instance vacancies in ``$d^{0}$'' compound as ZrO$_2$, HfO$_2$ or TiO$_2$.
By browsing the literature, one can find several examples which underline the importance and crucial role of the disorder. 
For example, in manganites one can observe after substitution, metal-insulator phase 
transition, formation of nanoscale inhomogeneities, or anomalies in the magnetic excitation spectrum, 
see for example \cite{Hennion,Karmakar,Ye,Coey-manganite}. We can also mention the high Curie temperature reported in materials which, a priori, do not contain any or at least a sufficient amount of magnetic impurities \cite{CoeyNature,Graphite1}. The ferromagnetism is attributed to defects which can in the case of HfO$_2$ appear close to the interface between the material and the substrate. Another example is the spin wave ferromagnetic excitation observed after vacancies are introduced in the cobaltate Na$_{1-x}\Box_{x}$CoO$_{2}$ \cite{Helme} (the symbol $\Box$ corresponds to vacancies), thus the substitution of Na by a vacancy is more than just a way to vary the carrier concentration. In diluted magnetic semiconductors as GaMnAs it is seen that the magnetic moment distribution and Curie temperature are also very sensitive to the sample history and thus to the presence of compensating defects as Mn interstitials and As anti-sites \cite{Kirby}.

In this manuscript we focus our attention on the nature of the magnetic excitation spectrum in dilute ferromagnetic materials. We will show that the proper and simultaneous treatment of both disorder and thermal fluctuations leads to unusual excitation spectrum in comparison to what is usually observed in non dilute systems. The paper is organized as follows. In the first part, we present a summary of the formalism based on the Self-Consistent Local Random Phase Approximation theory (SC-LRPA). In the second part, using the exchange couplings calculated within first principle approach (no adjustable parameters), we calculate the spectral function, excitation spectrum and spin stiffness as a function of the impurity concentration in 
{\it optimally annealed} Ga$_{1-x}$Mn$_{x}$As. The term {\it optimally annealed} means that the concentration of compensating defects is negligible. In the last part we compare our results to available experimental data for the spin stiffness and explain the large difference observed between as grown  and annealed samples.

The Hamiltonian which describes $N_{imp}$ interacting quantum (or classical) spins randomly distributed on a given lattice of N sites is the dilute Heisenberg model,
\begin{eqnarray}
 {\cal H}= - \sum_{{i},{j}} J_{ij} {\bf S}_{i} \cdot  {\bf S}_{j}
\end{eqnarray}

The couplings $J_{ij}$ are very general (short or long range), they are assumed to be given and can eventually depend on the impurity concentration $x=\frac{N_{imp}}{N}$. We will come back to this point in what follows.
For a given configuration of disorder e.g. the coordinates of the magnetic impurities, we define the retarded spin Green's function $G^{c}_{ij}(\omega)=\int_{-\infty}^{+\infty} G^{c}_{ij}(t)e^{i\omega t}dt$ where $ G^{c}_{ij}(t)= -i\theta(t)\langle [S_{i}^{+}(t),S_{j}^{-}(0)]\rangle$ which describes the transverse spin fluctuations, where the subscript $''c''$ corresponds to the configuration index.
After performing the Local Random Phase Approximation decoupling of the higher order Green's functions which appear in the equation of motion of $G^{c}_{ij}(\omega)$ \cite{EPL,PRB-compens}, we find,
\begin{eqnarray}
(\omega {\bf I} -{\bf H}^{c}_{\rm eff}) {\bf G}^{c} ={\bf D}
\label{eqp}
\end{eqnarray}
where ${\bf H}^{c}_{\rm eff}$, $G^{c}$ and ${\bf D}$ are $N_{imp} \times N_{imp}$ matrices. The index i runs over only the sites occupied by a localized spin. The effective Hamiltonian matrix elements is $({\bf H}^{c}_{\rm eff})_{ij} =  -\langle S_{i}^{z} \rangle J_{ij}+\delta_{ij} \sum_{l} \langle S_{l}^{z} \rangle J_{lj}$ and $D_{ij}= 2\langle S_{i}^{z} \rangle \delta_{ij}$. The local magnetization $\langle S_{i}^{z} \rangle$ has to be calculated {\it self-consistently} at each temperature. Note also that the condition $\sum_{j} ({\bf H}_{\rm eff}^{c})_{ij} =0$, implies that zero is eigenvalue of ${\bf H}^{c}_{\rm eff}$, the SC-LRPA treatment is consistent with the Goldstone theorem. 
 Note that although the matrix is non Hermitian, in the ferromagnetic phase the spectrum is real and positive at each temperature. If a negative eigenvalue would appear, this would indicate an instability of the ferromagnetic phase, as in the case where frustration would be present.
In the following, we will discuss the case of GaMnAs for which all the couplings are ferromagnetic, thus such instabilities will not occur.
Because ${\bf H}^{c}_{\rm eff}$ is non Hermitian (real non symetric), we now precise how the GF is expressed. ${\bf H}^{c}_{\rm eff}$, has the property to be bi-orthogonal \cite{Kuntz}. Thus, we define right and left eigenvectors of $H_{\rm eff}$ denoted respectively $|\Psi_{\alpha}^{R,c} \rangle $ and $|\Psi_{\alpha}^{L,c} \rangle $, both are associated to the same eigenvalues 
$\omega_{\alpha}^{c}$:  $ {\bf H}_{\rm eff}^{c} |\Psi_{\alpha}^{R,c} \rangle = \omega_{\alpha}^{c} |\Psi_{\alpha}^{R,c} \rangle$ and
 $ {\bf ^{t}H}_{\rm eff}^{c} |\Psi_{\alpha}^{L,c} \rangle = \omega_{\alpha}^{c} |\Psi_{\alpha}^{L,c} \rangle$.
In general, two eigenvectors belonging to the same set L or R are not orthogonal to each other, but when they are of different type they fulfill the relation $ \langle \Psi_{\alpha}^{R,c} | \Psi_{\alpha'}^{L,c} \rangle  = \delta_{\alpha,\alpha'}$. After inserting the L,R eigenvectors in eq.(\ref{eqp}), the retarded Green's function $G^{c}_{ij}(\omega)$ can be rewritten,
\begin{eqnarray}
G^{c}_{ij}(\omega)=\sum_{\alpha}\frac{ 2\langle S_{j}^{z} \rangle
}{\omega -\omega^{c}_{\alpha}+i\epsilon} \langle i  | \Psi^{R,c}_{\alpha}  \rangle \langle 
\Psi_{\alpha}^{L,c} | j \rangle 
\end{eqnarray}
Although the system is non translation invariant we define the Fourier transform by $\bar G({\bf q},\omega)= \langle \frac{1}{N_{imp}} \sum_{ij} e^{i{\bf q} (r_{i}-r_{j})}G^{c}_{ij}(\omega) \rangle_{c} $. The notation $\langle....\rangle_{c}$ denotes the average over the disorder configurations. We now define the dynamical spectral function $\bar A({\bf q},\omega)=\frac{-1}{\pi \langle \langle  S^{z}  \rangle \rangle} Im \bar G ({\bf q},\omega)$ where $\langle \langle  S^{z} \rangle \rangle = \frac{1}{N_{imp}}\sum_{i} \langle  S_{i}^{z}  \rangle $ is the average magnetization over the impurity sites. The spectral function $\bar A({\bf q},\omega)$ is the physical quantity which provides the direct access to the magnetic excitation spectrum, it is accessible by Inelastic Neutron Scattering experiment. It reads,
\begin{eqnarray}
\bar A({\bf q},\omega)=\langle \sum_{\alpha} A^{c}_{\alpha}({\bf q})\delta(\omega-\omega^{c}_{\alpha})\rangle_{c}
\end{eqnarray}
where $A^{c}_{\alpha}({\bf q})= \frac{1}{N_{imp}} \sum_{ij} \lambda_{j} \langle i | \Psi^{R,c}_{\alpha}  \rangle \langle \Psi^{L,c}_{\alpha} | j \rangle e^{i{\bf q} (r_{i}-r_{j})}$, we have introduced temperature dependent local parameter $\lambda_{j}= \frac{ \langle  S_{j}^{z}  \rangle}{\langle \langle  S^{z} \rangle \rangle }$. In the following we will calculate directly $\bar A({\bf q},\omega)$. However, It is also interesting to define the moments associated to it,
\begin{eqnarray}
m_{n}({\bf q})= \int_{-\infty}^{+\infty} \omega^{n} \bar A({\bf q},\omega) d\omega \nonumber \\
=\langle \frac{1}{N_{imp}}\sum_{ij}  
\sum_{\alpha} (\omega_{\alpha}^{c})^{n} \lambda_{j} \langle i | \Psi^{R,c}_{\alpha}  \rangle \langle \Psi^{L,c}_{\alpha} | j \rangle e^{i{\bf q} (r_{i}-r_{j})}\rangle_{c} 
\end{eqnarray}

We now make some general remarks. It is straightforward to see that the first moment can be rewritten $m_{1}({\bf q}) = \tilde{J}({\bf 0})-\tilde{J}({\bf q})$ where $\tilde{J}({\bf q}) = \langle \langle  S^{z}  \rangle \rangle \,\, \langle  \frac{1}{N_{imp}}\sum_{ij} \lambda_{j} \lambda_{i} J_{ij} e^{i{\bf q} (r_{i}-r_{j})}\rangle_{c}$. This expression is interesting and shows that the magnon excitation spectrum depends on the renormalized temperature dependent couplings by $J_{ij}^{\rm eff}(T) = J_{ij}  \lambda_{j} \lambda_{i}$. Since $\lambda_{i}$ are fluctuating from site to site, the previous expression indicates that the excitation spectrum will have a more complex temperature dependence than that obtained in the non dilute case ($x=1$ and $\lambda_{i}=1$) for which $\omega_{sw}(T)= \frac{\langle\langle S^{z} \rangle \rangle}{S} \,\, \omega_{sw}(T=0)$. In the limit of $q \rightarrow 0$, we immediately find that $m_{1}({\bf q}) \approx D_{1} q^{2}$ where the effective stiffness (in the following we will clarify the use of ``effective'') is $D_{1}=\langle \langle  S^{z} \rangle \rangle  \, \, \langle \frac{1}{2\,N_{imp}}\sum_{ij} \lambda_{j} \lambda_{i} J_{ij} ({\bf r}_{i} -{\bf r}_j)^{2}\rangle_{c} $. 

Note that, at low temperature where all spins are polarized, this expression can be simplified and does not depend on the nature of the eigenstates of the effective Hamiltonian.
If we assume that all couplings are ferromagnetic (as in GaMnAs) then we get, 
\begin{eqnarray}
D_{1}(x)= x D_{1}(x=1)
\end{eqnarray}
where $D_{1}= \frac{S}{2\,N} \sum_{ij} J_{ij} ({\bf r}_{i} -{\bf r}_j)^{2}$, where the double sum runs now over all sites.
This expression indicates that the spin stiffness extracted from the first moment calculation reduces only to the virtual crystal expression. The consequence of this is the following. First $D_{1}(x)$ overestimates the real spin stiffness. Additionally, at sufficiently low dilution, one expects that below the percolation threshold both $T_{C}$ and the spin stiffness should vanish. This expression is finite below the percolation threshold and does not know about it. Thus $D_{1}$ does not correspond to the real spin stiffness. In the following we will explain how to evaluate it properly.

The other interesting quantity is the effective linewidth $\gamma({\bf q})=\sqrt{m_{2}({\bf q})- m^{2}_{1}({\bf q}) }$, it measures the broadening of the magnetic excitations due to the disorder (dilution) and thermal fluctuations. In the long wave length limit, we can show that $\gamma({\bf q})= C q$. The linewidth is larger than the excitation energy. Similar results were reported in the study of the disordered double exchange model (nearest neighbor coupling) relevant for manganite compounds ($x=1$) \cite{Motome}. Then, one would naively be tempted to conclude that there is no well defined magnetic excitations. As it will be seen in the following, even though the magnon excitations are incoherent, they appear as well defined peak in $A({\bf q},\omega)$. In other words, the energy and width of the excitations calculated with the moments of the spectral function do not correspond to the peak position and linewidth of the excitations directly calculated with the full spectral function.

\begin{figure}[tbp]
\includegraphics[width=6.cm,angle=-90]{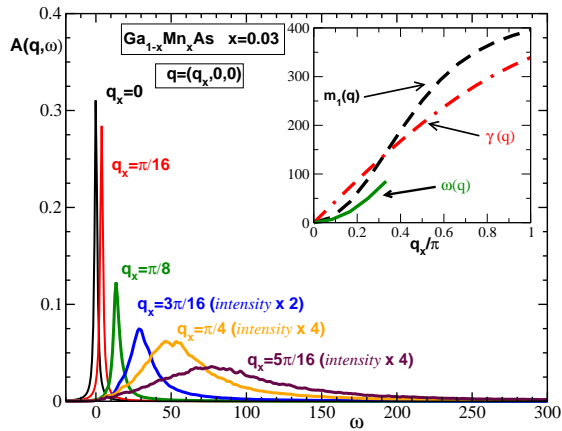}
\caption{(Color online)
Spectral function for Ga$_{1-x}$Mn$_{x}$As as a function of the energy $\omega$ (in Kelvin) in the $(1,0,0)$ direction for different values of $q_{x}$, with $x=0.03$. In the inset, are plotted the first moment, the effective linewidth and the position of the peak in $A(q,\omega)$ as a function of $q_{x}/\pi$(in units of $a^{-1}_{0}$) and the y-axis is in Kelvin.}
\label{sqw}
\end{figure}

In the following section and to illustrate our theory we calculate the magnetic excitation spectrum in the III-V dilute magnetic semiconductor Ga$_{1-x}$Mn$_{x}$As as a function of the Mn$^{2+}$ concentration and at $T= 0 ~K$. In this case, $\lambda_{i}=1$, the matrix $H_{\rm eff}$ is real symetric and thus L and R eigenvectors are identical. We present a detailed analysis of the excitation spectrum as a function of the $Mn^{2+}$ concentration. In addition, in order to allow a quantitative study with no fitting parameters and a direct comparison with the experimental measurements, the magnetic exchange coupling we use were calculated from first principle Tight Binding Linear Muffin Tin Orbital approach. The same couplings were used successfully for example in ref.\cite{EPL,PRB-compens} to calculate the Curie temperature. We stress that the used exchange couplings extend over 62 shells.

In Fig.\ref{sqw} we have plotted $\bar A(q,\omega)$ for Ga$_{1-x}$Mn$_{x}$As as a function of energy for different values of the momentum {\bf q} in the (1,0,0) direction. We remind that GaAs has a zinc blend structure, thus the magnetic impurities are distributed randomly on the fcc sublattice of Ga. In addition, periodic boundary conditions have been used. The calculations were performed on a system which contains $ 4\times (32)^{3}$ sites, the average over disorder was performed over 50 configurations. Note that for $x=0.03$ the system contains approximately $N_{imp}=4000$ impurities. In all cases, we have checked that the number of disorder configurations was sufficient to provide accurate results.Indeed, for example for the $ 4\times (16)^{3}$ system, it was found that the average over 50 configurations is already sufficient. Even if we increase the number of configurations $\bar A(q,\omega)$ remains almost unaffected.

 In Fig.\ref{sqw} we can observe well defined excitations for only relatively small values of the momentum. As the momentum q increases the peak becomes broader and the well defined excitation is then replaced by a very broad structure. In the inset we have plotted the first moment $m_{1} (q)$, the linewidth $\gamma(q)$ and the position of the peak which we denote as $\omega(q)$ as a function of the momentum $q$. It is clearly seen that at low $q$ both $m_{1}(q) \approx D_{1}q^{2}$ and $\omega(q) \approx Dq^{2}$. However the ``effective stiffness'' $D_{1}$ appears to be larger than D, typically in this case we find $D_{1}/D \approx 2 $. This is in agreement with the previous discussion that $D_{1}$ overestimates the real spin stiffness, especially as we approach the regime of percolation. The main consequence of this is that the accurate determination of the spin stiffness should be done by directly plotting the energy of the peak as a function of the momentum instead of using the first moment dispersion. As mentioned in the previous section we indeed observe that the effective linewidth $\gamma(q)$ is linear in q, but it does not correspond to the real linewidth of the peak observed in $\bar A(q,\omega)$. The second moment is inappropriate to 
evaluate the real linewidth of the magnetic excitation, this is reflected by the fact that the peak observed in Fig.\ref{sqw} are non Gaussian-like and very asymmetric. In fact, asymmetric peaks in the magnetic excitation spectrum were observed in the diluted antiferromagnet Mn$_{x}$Zn$_{1-x}$F$_{2}$ \cite{Uemura-Birgeneau}. It was observed that as we approach the percolation threshold $x_{c}=0.25$ the line shape of the low magnetic excitation peaks becomes strongly asymmetric and develops a long tail extending towards higher energy. This is clearly seen in Fig.\ref{sqw}. It was shown that the peaks can be fitted by a two component structure: a sharp Gaussian peak which describe the magnon excitation (its energy is q-dependent) and a broad localized mode (weakly q dependent). The observed increase of asymmetry of the peak as we increase the momentum corresponds to a cross over from propagating low energy excitations to localized excitations (fractons) \cite{Orbach,Aharony}.

\begin{figure}[tbp]
\includegraphics[width=7cm,angle=-90]{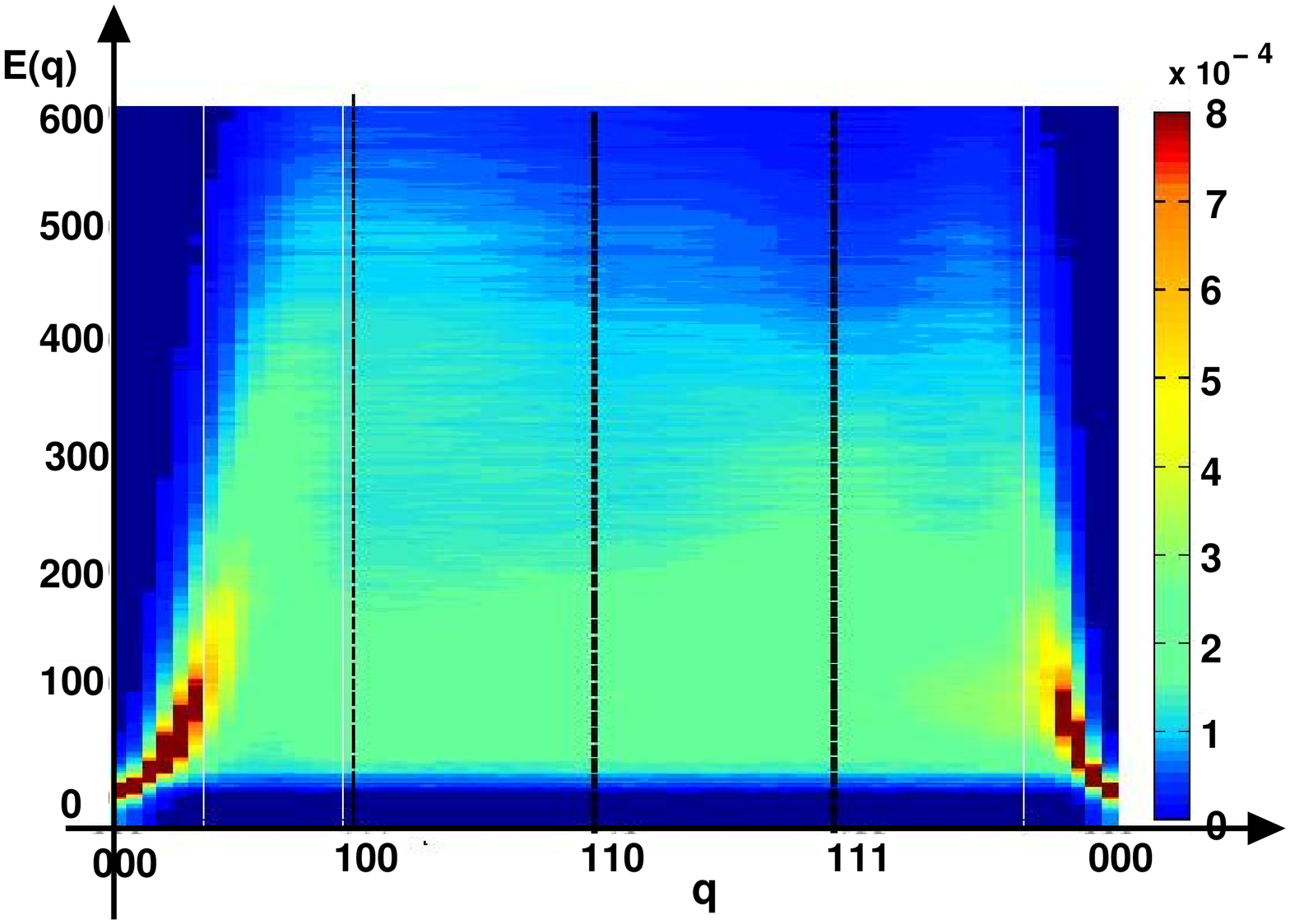}
\caption{(Color online) 
Spectral function $\bar A(q,\omega)$ in Ga$_{1-x}$Mn$_{x}$As in the $(q,\omega)$ plane ($x=0.03$). 
The energy axis (y axis) is in Kelvin
}
\label{sqw-color}
\end{figure}

In the next figure, Fig.\ref{sqw-color}, we have plotted the spectral function in $(q,\omega)$ plane. As in the previous figure the density of magnetic impurity is $x=0.03$. In contrast to what is usually observed in weakly disordered magnetic systems as in manganites well defined excitations exist in the dilute case only in a restricted region of the Brillouin zone centered around the $\Gamma$ point ($q=(0,0,0)$).
This unusual feature shows that the nature of ferromagnetism in the diluted semiconductors is very different from that observed in non dilute materials. As seen in the previous figure the broadening of the excitation increases significantly as we move away from the $\Gamma$ point. We expect that the momentum cut-off below which a well defined excitation exists should be related to the percolation correlation length $\xi_{p}$ \cite{Uemura-Birgeneau}.

\begin{figure}[tbp]
\includegraphics[width=7cm,angle=-90]{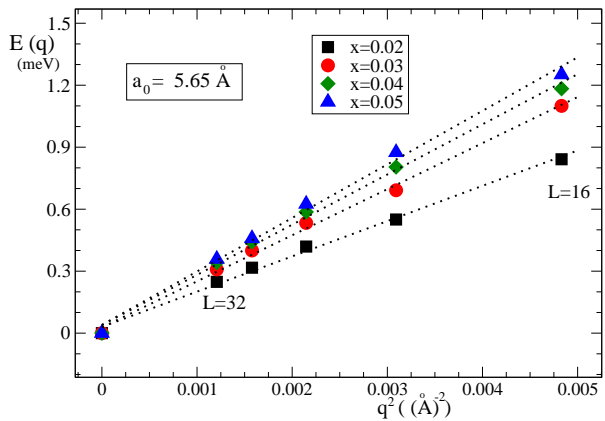}
\caption{(Color online)
Magnon energy $\omega(q)$ in meV as a function of $q^{2}$ (in $\AA^{-2}$) for different concentration of $Mn^{2+}$ impurities ($x=0.02$,0.03,0.04 and 0.05). The size L of the fcc lattice varies from $L=16$ to $L=32$.}
\label{eqfq2}
\end{figure}

\begin{figure}[tbp]
\includegraphics[width=6cm,angle=-90]{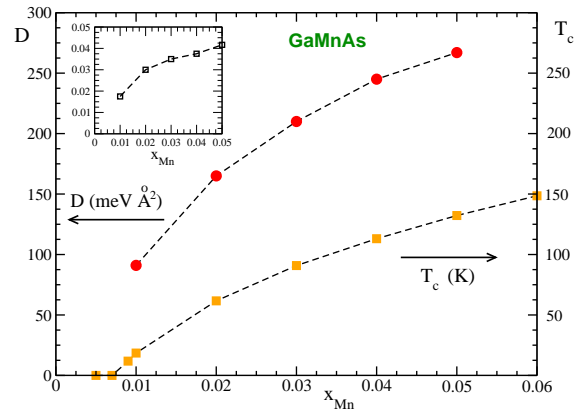}
\caption{(Color online) Spin stiffness D in $meV \AA^{2}$ as a function of the concentration x in optimally annealed samples and $T_{C}$ in Kelvin in Ga$_{1-x}$Mn$_{x}$As from ref. \cite{EPL}. In the inset the ratio $T_{C}/D$ (in $\AA^{-2}$)as a function of x is plotted.
}
\label{stiffness}
\end{figure}

In Fig.\ref{eqfq2} we have plotted, for different concentration of Mn impurities, the spin wave energy $\omega(q)$ at low q. To be more precise the energy of the first peak in $A(q,\omega)$ in the (1,0,0) (${\bf q}= (\frac{2\pi}{L a_{0}},0,0)$ is shown as a function of $q^{2}$, the system size varies from $L=16$ to $L=32$. We 
clearly observe that for small momentum q $\omega (q) \propto D(x) q^{2}$.
Note that we have used for $a_{0}=5.65 \AA$ the value of the lattice spacing in GaAs.Note that, the error bars are included in the size of the symbols. If the average over disorder configuration was not sufficient one would expect that the curve $\omega (q)$ would exhibit some noise. In Fig.\ref{stiffness}, we have plotted the spin stiffness D(x) and the Curie temperature (from ref.\cite{PRB-compens}) as a function of the $Mn^{2+}$ concentration x. We remind that there is no adjustable parameters in these calculations. We observe that the values of $D(x)$ are relatively high for dilute ferromagnetic system. For example for $x=0.03$ for which the Curie temperature is $T_{C} \approx 80 \,K$ the spin stiffness is $D \approx 210 \,meV \AA^{2}$. Interestingly this value is close to what is usually measured in non dilute systems as in manganites  \cite{Martin,Fernandez,Endoh}.
When approaching the percolation threshold from above the spin stiffness decreases very fastly, faster than the Curie temperature which can be fitted by $ T_{C} \approx A(x-x_{c})^{1/2}$ \cite{PRB-compens}. However we can not extract the critical power in a reliable manner, this should require more calculations close to the percolation threshold. In the inset we have also plotted the ratio $T_{c}/D$, it is clearly seen that the Curie temperature is not proportional to the spin stiffness.

It is now interesting to compare these results to experimental data. In contrast to Curie temperature or resistivity measurements, there are only few experimental studies concerning the magnetic excitation in diluted magnetic semiconductors. In our knowledge there are two papers in which the measurement of the spin stiffness in GaMnAs was done. In ref.\cite{Goennenwein}, the authors have measured it by ferromagnetic resonance in as grown sample where $x=0.05$. They have reported a value of approximately $60 \,meV \,\AA^{2}$.
In a recent study \cite{Wang}, done using samples of different concentrations and sample history, it was shown that annealed sample possess much higher values of the spin stiffness.
Indeed, it was shown that the annealed sample ($x=0.03$) which orders ferromagnetically below $T^{exp}_{C}= \,80 K$ has a spin stiffness of the order of $D^{exp} \approx 160\, meV \AA^{2}$, whilst for as grown samples ($x=0.05$ and $0.06$) $D^{exp}$ was only approximately $40 \, meV \AA^{2}$.
We remind that our calculation are performed for optimally annealed samples. In  Fig.\ref{stiffness} for $x=0.03$ we find $T_{C}^{th} = 88 \,K$ and $D^{th}(x=0.03) = 210  \,meV \AA^{2}$, both of these values agrees well with the experimental values. Because the concentration of Mn is not precisely known in experimental samples, a value of $x=2.5 \%$ would lead to $T_{C}^{th} =75 K$ and $D^{th}(x=0.025) = 175  \, meV \AA^{2}$, this would significantly improve the agreement. Now the interesting question is: Why are the values of the spin stiffness is as grown sample so small?
Let us assume that before annealing the main effect of compensation (for more details see ref.\cite{PRB-compens}) is a reduction of the density of the magnetically active Mn ($x_{\rm eff}$). Because the measured $T_{C} \approx 65-80 \, K$ for $x \approx 5\% - 6\%$ , from the theoretical curve of $T_{C}$ we would conclude that $x_{\rm eff} \ge 2.5 \%$. Thus the expected stiffness should be larger than 150 $\, meV \AA^{2}$. This is in disagreement with the much smaller experimental values. We argue that the physical mechanism which appears to be responsible for such small values should be the inhomogeneous nature of the as grown sample instead of the reduction of the magnetically active localized spins. In ref. \cite{PRB-compens}, we could explain the measured $T_{C}$ in as grown sample by just estimating for each of them $x_{\rm eff}$ but with no need to take into account of the inhomogeneous nature. Thus D is more sensitive to the presence of inhomogeneities than the Curie temperature \cite{Bouzerarnew}. 

To conclude, we have calculated the magnetic excitation spectrum in dilute ferromagnets within a theory based on the Self-consistent Local RPA. The theory is general and allows to treat disorder and thermal fluctuations in a reliable manner. As an illustration, we have calculated the spectrum in well annealed sample of GaMnAs as a function of the impurity concentration. It is predicted that well defined excitation can be observed only in a restricted region of the momentum space centered around the $\Gamma$ point. A good agreement with experimental values of the spin stiffness was obtained for the annealed sample with concentration $x=0.03$. We have also discussed and provided an explanation to  the very small value of the spin stiffness measured in as grown samples. We hope that this theoretical study will motivate new experimental measurements for example D as a function of the impurity concentration and as a function of the annealing treatment.

I would like to thank E. Kats, B. Barbara, J. Kudrnovsky, P. Nozi\`eres for useful discussions,
R. Bouzerar and O. Cepas for carefully reading the manuscript and E. Kats for the hospitality 
at the Laue Langevin Institut and providing computer facilities.

\end{document}